\documentclass{article} 
\usepackage{iclr2025_conference,times}

\usepackage{amsmath,amsfonts,bm}

\def\eqref#1{equation~\ref{#1}}

\def\1{\bm{1}}

\DeclareMathAlphabet{\mathsfit}{\encodingdefault}{\sfdefault}{m}{sl}
\SetMathAlphabet{\mathsfit}{bold}{\encodingdefault}{\sfdefault}{bx}{n}

\usepackage{hyperref}
\usepackage{url}
\usepackage{graphicx}
\usepackage{enumitem}

\title{Watermarking for AI Content Detection: A Review on Text, Visual, and Audio Modalities}

\author{Lele Cao \\
AI Labs, King / Microsoft\\
Stockholm, Sweden \\
\texttt{lele.cao@king.com}\\ 
\texttt{lelecao@microsoft.com}
}

\iclrfinalcopy 
\begin{document}

\maketitle

\begin{abstract}
The rapid advancement of generative artificial intelligence (GenAI) has revolutionized content creation across text, visual, and audio domains, simultaneously introducing significant risks such as misinformation, identity fraud, and content manipulation. This paper presents a practical survey of watermarking techniques designed to proactively detect GenAI content. We develop a structured taxonomy categorizing watermarking methods for text, visual, and audio modalities and critically evaluate existing approaches based on their effectiveness, robustness, and practicality. Additionally, we identify key challenges, including resistance to adversarial attacks, lack of standardization across different content types, and ethical considerations related to privacy and content ownership. Finally, we discuss potential future research directions aimed at enhancing watermarking strategies to ensure content authenticity and trustworthiness. This survey serves as a foundational resource for researchers and practitioners seeking to understand and advance watermarking techniques for AI-generated content detection.
\end{abstract}

\section{Introduction}

The exponential growth of generative artificial intelligence (GenAI) has reshaped the digital landscape, enabling the automated creation of realistic text, images, videos and audio. State-of-the-art generative models, such as DeepSeek \citep{guo2025deepseek}, GPT series \citep{Radford2018Improving}, Stable Diffusion \citep{esser2024scaling}, DALL·E \citep{ramesh2021zero}, MusicGen \citep{NEURIPS2023_94b472a1}, and NaturalSpeech \citep{tan2024naturalspeech} have demonstrated remarkable capabilities in mimicking human creativity. While these advancements show transformative opportunities in media, entertainment, and communication, they also pose substantial risks, including misinformation, identity fraud, and content manipulation \citep{chen2024demamba,tang2024science}. The ability of GenAI content to seamlessly blend into real-world scenario raises concerns on content authenticity and trustworthiness.

To counteract these risks, AI-generated content detection has emerged as a critical research domain. Among various detection methods, watermarking stands out as a proactive solution that embeds imperceptible yet traceable signatures within AI-generated outputs. Unlike reactive detection approaches that rely on post-generation analysis, watermarking enables content origin verification at the point of creation, facilitating robust and scalable AI-content identification \citep{lee2023wrote}. This makes watermarking particularly attractive for combating the misuse of AI-generated media.

Watermarking techniques vary significantly across different content modalities. For text-based AI content, probabilistic watermarking methods adjust token distributions to introduce hidden patterns, making it possible to detect GenAI text without impacting readability \citep{kirchenbauer2023watermark}. In visual content, watermarking strategies include spatial-domain embedding, frequency-domain modifications, and deep learning-based approaches to ensure robust detection against adversarial alterations \citep{mavali2024fake}. For GenAI audio, watermarking methods modify spectral and temporal properties, allowing the authentication of synthetic speech and music \citep{roman:hal-04610152}.

Despite the promise of watermarking, several challenges remain. Watermark robustness is a key issue, as adversarial attacks and transformation techniques can be used to remove or obfuscate watermarks \citep{sadasivan2023can}. Moreover, the standardization of watermarking practices across modalities remains an open research problem, with different industries adopting varying approaches to embedding and detection. Addressing these challenges is essential for the widespread adoption of watermarking in AI-generated content detection.

This survey aims to provide a comprehensive analysis of watermarking techniques across multiple AI-generated content modalities. Specifically, we
\begin{itemize}[topsep=0pt, partopsep=0pt, itemsep=0pt]
    \item Present a structured taxonomy of watermarking methods used in AI-generated text, visual, and audio content detection.
    \item Review and compare existing watermarking techniques in terms of effectiveness, robustness, and practicality.
    \item Identify critical challenges in watermarking AI-generated content and discuss potential future research directions.
\end{itemize}

It is important to note that previous surveys have provided valuable insights into GenAI content detection for specific modalities, such as textual \citep{crothers2023machine,tang2024science,valiaiev2024detection,yang2024survey,chaka2024reviewing,wu2025survey}, visual \citep{passos2024review,deng2024survey,sandotra2024comprehensive}, and audio content \citep{almutairi2022review,yi2023audio,li2024audio,pham2024comprehensive,li2024audio}. However, none of these studies have offered a unified treatment of all three modalities. Furthermore, even two recent surveys that cover multimodal AI-generated content detection \citep{lin2024detecting,yu2024fake} have only briefly discussed watermarking as one of several detection approaches. In contrast, our survey is the first to systematically review watermarking techniques across text, visual, and audio domains. This integrated approach not only highlights the common underlying principles but also addresses the unique challenges and opportunities specific to each modality, thereby offering a holistic perspective that bridges existing gaps in the literature.

The remainder of this paper is structured as follows: Section~\ref{sec_wm_fundamental} introduces the fundamental principles of watermarking and key evaluation metrics. Sections \ref{sec_wm_text}, \ref{sec_wm_vis}, and \ref{sec_wm_audio} explore watermarking techniques for text, visual, and audio modalities, respectively. Section~\ref{sec_challenge_ethical_future} discusses open challenges, ethical considerations and future directions, followed by a conclusion in Section~\ref{sec_conclusion}.

\section{Fundamentals of Watermarking}
\label{sec_wm_fundamental}

Watermarking is a proactive technique designed to embed imperceptible yet identifiable markers within AI-generated content, facilitating its detection and authentication. Unlike passive detection methods that analyze content post-generation, watermarking introduces traceable signatures at the point of content creation, enabling robust verification and origin attribution. This section formally defines watermarking schemes, explores different types of watermarking approaches, and discusses key evaluation metrics.

\subsection{Definition and Key Principles of Watermarking}
\label{sec:def}

\begin{figure}[t!]
    \centering
    \includegraphics[width=\linewidth]{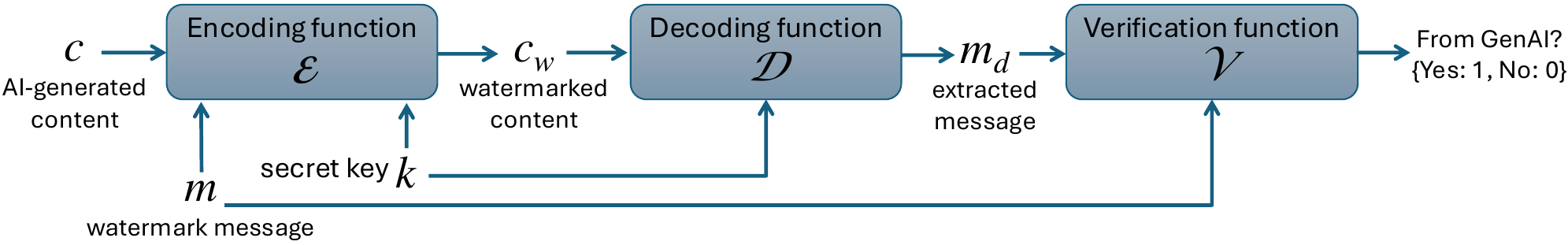}
    \caption{Schematic watermarking procedure: verify if input content $c$ is produced from GenAI or a specific GenAI model.}
    \label{fig:watermarking-procedure}
\end{figure}

A watermarking scheme, visualized in Figure~\ref{fig:watermarking-procedure}, for AI-generated content can be defined as a tuple:
\begin{equation*}
\label{eq_def}
\mathcal{W} = (\mathcal{E}, \mathcal{D}, \mathcal{V}),\, \text{where}
\end{equation*}
\begin{itemize}[leftmargin=10pt, topsep=0pt, partopsep=0pt, itemsep=0pt]
\item $\mathcal{E}: \mathcal{C} \times \mathcal{K} \times \mathcal{M} \to \mathcal{C}_w$ is the encoding function that embeds a watermark message $m \in \mathcal{M}$ into the input content $c \in \mathcal{C}$ using a secret key $k \in \mathcal{K}$, producing a watermarked output $c_w \in \mathcal{C}_w$.
The set $\mathcal{C}$ contains original content, which may include texts, images, or audio signals.
$\mathcal{C}_w \subset \mathcal{C}$ is the set of watermarked content.
The sets $\mathcal{K}$ and $\mathcal{M}$ jointly guide the encoding and decoding processes. Depending on the modality, watermarks can take different forms: a text watermark might appear as a specific distribution of words (cf.~chp1 in \citealp{cao2025practical}), a visual watermark could be a textual message (cf.~chp2 in \citealp{cao2025practical}), and an audio watermark may be embedded as an additive waveform (cf.~chp3 in \citealp{cao2025practical}).

\item $\mathcal{D}: \mathcal{C}_w \times \mathcal{K} \to \mathcal{M}$ is the decoding function that attempts to extract the watermark message $m_d$ from the watermarked content $c_w$ using the key $k$.
\item $\mathcal{V}: \mathcal{M} \times \mathcal{M} \to {0,1}$ is the verification function that determines whether the extracted watermark message $m_d$ is a valid watermark.
\item $\mathcal{C}$ is the set of original content, which may include texts, images, or audio signals.
\end{itemize}

A robust watermarking scheme should exhibit the following properties:
\begin{itemize}[leftmargin=10pt, topsep=0pt, partopsep=0pt, itemsep=0pt]
\item \textbf{Imperceptibility:} The watermark should not degrade content quality or be detectable through normal human perception \citep{kirchenbauer2023watermark}. Visual and auditory contents are evaluated with PSNR (peak signal-to-noise ratio). For text, BLEU (bilingual evaluation understudy) and ROUGE (recall-oriented understudy for gisting evaluation) scores measure similarity between original and watermarked text.
\item \textbf{Robustness:} The watermark should remain intact even after transformations such as compression, cropping, paraphrasing, or noise injection \citep{lee2023wrote}. This is evaluated through BER (bit error rate) defined as $\frac{1}{N} \sum_{i=1}^{N} \mathbb{1}(m_i \neq \hat{m}_i)$, where $m_i$ and $\hat{m}_i$ are the original and extracted watermark bits, respectively. Lower BER indicates greater robustness \citep{roman:hal-04610152}.
\item \textbf{Security:} The watermark should be resistant to unauthorized removal or forgery by adversarial attacks. Watermarking security is often assessed based on its resistance to adversarial attacks such as GAN-based watermark removal and synonym substitution \citep{sadasivan2023can}. Techniques such as key-based authentication and encrypted watermarking enhance security.
\item \textbf{Capacity:} The scheme should allow embedding sufficient information without significantly altering the content \citep{li2024revisiting}.
\end{itemize}
Interested readers are referred to papers like \citep{aberna2024digital} for a more comprehensive side-by-side comparison of various watermarking performance metrics.

\subsection{Types of Watermarking}

Watermarking techniques vary based on their embedding methodology, visibility, and robustness to modifications. These techniques can be categorized into several types.

\textbf{Visible vs. invisible watermarking.} Visible watermarking embeds explicit identifiers such as logos or text overlays, commonly used in visual media but impractical for text and audio. Invisible watermarking, on the other hand, introduces imperceptible modifications that require algorithmic analysis for detection \citep{sharma2024review}.

\textbf{Fragile vs. robust watermarking.} Fragile watermarking is sensitive to modifications and is primarily used for content integrity verification, while robust watermarking remains detectable even after transformations such as compression, cropping, or reformatting.

\textbf{Spatial vs. frequency domain watermarking.} Spatial domain watermarking directly modifies content features, such as altering pixel values in images or inserting specific words in text. Frequency domain watermarking embeds information in transformed representations, such as using Discrete Cosine Transform (DCT) \citep{PARAH201611} or Discrete Wavelet Transform (DWT) \citep{hurrah2019dual} to modify frequency coefficients.

\textbf{Deep learning-based watermarking.} Recent advancements leverage deep neural networks to encode watermarks into the latent representations of AI-generated content, improving imperceptibility and robustness.

\section{Watermarking for Text-Based AI Content Detection}
\label{sec_wm_text}

The proliferation of large language models (LLMs) such as GPT-4~\citep{achiam2023gpt}, Llama-3~\citep{dubey2024llama} and DeepSeek~\citep{guo2025deepseek} has revolutionized text generation, enabling the production of human-like content across various domains. While these advancements enhance accessibility and efficiency, they also introduce risks related to misinformation, academic dishonesty, and content manipulation. The challenge of distinguishing AI-generated text from human-authored content has spurred the development of watermarking techniques, which embed traceable markers within AI-generated text to facilitate its detection. Unlike post-hoc detection methods that analyze textual artifacts, watermarking offers a proactive solution by encoding verifiable signatures at the point of generation, allowing for robust attribution and authentication.

\subsection{Principles of Text Watermarking}

Watermarking for text-based AI content detection operates by embedding hidden yet detectable patterns into LLM-generated text. Formally, a text watermarking scheme consists of an encoding function that modifies generated tokens according to a predefined scheme, a decoding function that extracts and verifies the watermark, and a verification function that determines the presence of a valid watermark. These techniques must balance imperceptibility, robustness, security, and capacity to ensure effectiveness.

Imperceptibility ensures that watermarked text remains natural and indistinguishable from non-watermarked text to human readers. Robustness is necessary to maintain watermark detectability even after paraphrasing, synonym substitution, or minor modifications. Security prevents adversarial attacks from removing or falsifying watermarks. Capacity determines how much information can be embedded within a given text length without significantly altering its meaning or coherence.

\subsection{Probabilistic Token-Level Watermarking}

A widely adopted approach for text watermarking involves manipulating token selection probabilities during text generation. \cite{kirchenbauer2023watermark} proposed a probabilistic watermarking scheme that biases the token sampling process toward a subset of “green” tokens based on a secret key. During generation, at each token step, the LLM preferentially selects tokens from this predefined green list, subtly altering the probability distribution. The resulting text appears natural but contains statistical deviations that are detectable by an authorized verifier.

Watermark detection relies on analyzing token distributions in a given text sample. If the proportion of green tokens exceeds a statistically significant threshold, the text is classified as watermarked. This method offers strong robustness against paraphrasing attacks while maintaining readability. However, challenges arise when texts are heavily edited, as modifications may disrupt the statistical signal, reducing detectability.

\subsection{Lexical and Syntactic Watermarking}

Beyond probabilistic methods, lexical and syntactic watermarking techniques modify word choices and sentence structures in a controlled manner. Early lexical watermarking techniques involved embedding predefined sequences of rare words or phrases within AI-generated text. However, these methods are susceptible to removal through simple editing \citep{roman:hal-04610152}. More sophisticated approaches leverage synonym substitution constrained by a secret embedding space, ensuring that word replacements are semantically valid while encoding a detectable pattern \citep{DBLP:conf/iclr/LiuPHM024}.

Syntactic watermarking operates by controlling sentence structures or punctuation patterns. A method proposed by \cite{lee2023wrote} involves encoding information into the grammatical structure of sentences without altering meaning. For instance, an AI-generated sentence might consistently favor passive constructions or specific syntactic tree patterns. These patterns are statistically rare in human-authored text, making them detectable while preserving naturalness.

\subsection{Contextual and Semantic Watermarking}

Recent advancements explore contextual and semantic watermarking, embedding hidden signals into higher-level linguistic structures. One approach involves strategically inserting information-bearing paraphrases or stylistic markers into text, which remain robust against superficial modifications. For example, a watermarking scheme might enforce subtle stylistic consistencies, such as recurring phrase structures, discourse markers, or thematic redundancies \citep{sadasivan2023can}.

Another promising direction is embedding watermarks in semantic embeddings rather than raw text. Instead of altering words directly, this technique adjusts the latent representations from which text is generated, ensuring that semantic consistency is maintained while encoding verifiable signals \citep{roman:hal-04610152}. Such methods provide robustness against paraphrasing attacks but require specialized decoding mechanisms to extract watermark signatures from transformed text.

\section{Watermarking for AI-Generated Visual Content}
\label{sec_wm_vis}

The rise of AI-generated visual content has introduced transformative applications across media, design, and entertainment, yet it also raises substantial concerns regarding misinformation, copyright infringement, and digital authenticity. Advances in generative models, particularly generative adversarial networks (GANs) \citep{goodfellow2014generative}, diffusion models \citep{ho2020denoising}, and variational autoencoders (VAEs) \citep{Kingma2014}, have enabled the creation of hyper-realistic images and videos that often evade human scrutiny. The necessity of robust detection mechanisms has thus led to extensive research on watermarking techniques designed to embed traceable signatures within AI-generated visual content \citep{saberi2024robustness}. Unlike reactive detection approaches, watermarking offers a proactive solution by embedding identifying marks directly into images and videos at the point of generation, facilitating verification and attribution.

Watermarking techniques for visual content can be categorized based on embedding methodologies, resilience against transformations, and their detectability. Broadly, these approaches include spatial-domain embedding, frequency-domain transformations, hybrid methods, and deep learning-based watermarking. Each of these techniques presents unique strengths and challenges in terms of imperceptibility, robustness, and resistance to adversarial attacks \citep{hosny2024digital}.

\subsection{Spatial-Domain Watermarking}

Spatial-domain watermarking directly modifies pixel values to encode information, making it straightforward to implement and computationally efficient. Traditional methods such as Least Significant Bit (LSB) embedding insert watermarks into the least significant bits of selected pixels, ensuring minimal visual distortion \citep{wolfgang1996watermark}. However, this approach is highly vulnerable to common image processing operations such as compression, filtering, and noise addition, which can degrade or remove the watermark \citep{sharma2024review}.

To enhance robustness, spatial-domain watermarking techniques have evolved to incorporate more sophisticated embedding strategies. Local Binary Pattern (LBP)-based watermarking improves resilience by encoding patterns in texture descriptors rather than direct pixel modifications \citep{wenyin2011semi}. Similarly, cryptographic hashing techniques combined with spatial embedding increase security by ensuring that the watermark is difficult to tamper with without knowledge of the key \citep{raj2018blockwise}. Despite these advancements, spatial-domain watermarking remains sensitive to geometric distortions such as cropping and resizing, limiting its reliability in practical applications.

\subsection{Frequency-Domain Watermarking}

Frequency-domain watermarking addresses the vulnerabilities of spatial-domain methods by embedding watermarks within transformed representations of an image, making them more resistant to compression and noise \citep{singh2024digital,sharma2024review}. The most common frequency-domain transformations used for watermarking include the Discrete Cosine Transform (DCT), Discrete Wavelet Transform (DWT), and Singular Value Decomposition (SVD).

DCT-based watermarking embeds information in mid-frequency coefficients, ensuring robustness against compression while maintaining visual quality \citep{PARAH201611}. However, its reliance on fixed-sized blocks can introduce artifacts that affect imperceptibility. DWT-based approaches offer improved resistance to geometric transformations by embedding watermarks across multiple frequency bands, balancing robustness and visibility \citep{hurrah2019dual}. SVD-based watermarking enhances security by modifying singular values, which remain stable under common transformations like compression and noise addition \citep{benrhouma2017security}. Nevertheless, these methods can suffer from synchronization issues when faced with significant image distortions or adversarial modifications \citep{luo2023dvmark}.

\subsection{Hybrid Watermarking Techniques}

Recent advancements, such as DWT-DCT-SVD watermarking \citep{kang2018novel}, combine multiple transform domain methods or integrate spatial domain techniques to boost robustness, flexibility, and real-time performance. The hybrid approaches try to embed multiple watermarks to address various scenarios, such as high visibility in one layer and resilience in another. The hybrid technique is especially useful for applications requiring protection against a combination of attacks, such as adversarial or spoofing attacks, and is resilient to common image transformations like compression and noise addition \citep{haghighi2021wsmn}.

\subsection{Deep Learning-Based Watermarking}

Deep learning-based watermarking represents a significant evolution in AI-generated visual content detection, offering adaptive and highly resilient techniques. These methods leverage neural networks to encode watermarks into latent representations, ensuring robust embedding while preserving visual fidelity \citep{singh2024digital}. Two prominent examples of deep learning-driven watermarking solutions are Google's SynthID\footnote{SynthID: \url{https://deepmind.google/technologies/synthid}.} and Meta’s Stable Signature \citep{fernandez2023stable}.

SynthID embeds imperceptible watermarks into GenAI images, allowing detection even after transformations like cropping and filtering. Unlike traditional methods, SynthID operates in the latent space of image generation models, ensuring watermarks remain intact even after extensive modifications. Stable Signature follows a similar principle but integrates watermarking directly in the diffusion process, embedding unique binary signatures that persist under common transformations.

These deep learning approaches offer substantial advantages in terms of robustness and adaptability. However, they also present challenges, particularly in terms of computational complexity and susceptibility to adversarial attacks. Recent studies have shown that diffusion-based models can be fine-tuned to erase watermarks while maintaining visual consistency, highlighting an ongoing arms race between watermarking techniques and adversarial removal strategies~\citep{saberi2023robustness}.

\section{Watermarking for AI-Generated Audio Content}
\label{sec_wm_audio}

The fast surge of AI-generated audio, encompassing both synthetic speech and AI-composed music, has introduced significant challenges related to authenticity verification, copyright protection, and deepfake detection. Advanced generative models, such as text-to-speech (TTS) systems and AI-driven music composition platforms, have achieved remarkable realism, making it increasingly difficult to differentiate between synthetic and human-generated audio \citep{li2024audio,li2024detecting}. While detection methods based on acoustic feature analysis and deep learning classifiers provide post-hoc identification of AI-generated content, watermarking presents a proactive solution that embeds identifiable, yet imperceptible, markers within AI-generated audio at the point of creation. This section provides a comprehensive review of watermarking techniques for AI-generated speech and music, discussing fundamental methodologies, implementation strategies, and ongoing challenges in robustness and adversarial resilience.

\subsection{Principles of Audio Watermarking}

Audio watermarking operates by embedding an imperceptible yet verifiable signal within an audio waveform, ensuring the traceability and integrity of AI-generated speech and music. A watermarking scheme for audio content follows the standard tuple formulation in Section~\ref{sec:def}: an encoding function that embeds the watermark, a decoding function that extracts it, and a verification function that authenticates its presence. Again, these schemes must meet four primary requirements: 
(1) imperceptibility requires embedding techniques that do not degrade audio quality; (2) robustness ensures the watermark survives transformations such as compression, noise addition and remixing; (3) security protects against unauthorized removal or spoofing; and (4) capacity defines how much information can be embedded without compromising detectability.

Watermarking approaches for AI-generated audio can be categorized into two primary domains: speech watermarking, which focuses on embedding watermarks in synthetic voices, and music watermarking, which targets AI-composed musical compositions. The following subsections examine each category in detail, reviewing major methodologies and their comparative advantages.

\subsection{Watermarking for AI-Generated Speech}

The advent of high-fidelity TTS and voice conversion (VC) models has enabled the creation of hyper-realistic synthetic voices, raising concerns about voice deepfakes and identity fraud. Watermarking AI-generated speech provides a reliable method for verifying audio authenticity, embedding inaudible markers that can be extracted and validated post-generation. Speech watermarking techniques are broadly categorized into spectral/temporal-domain and deep learning-based approaches.

\textbf{Spectral-domain watermarking} embeds imperceptible markers within the frequency components of an audio signal, leveraging transformations such as the Discrete Fourier Transform (DFT), DCT, and DWT \citep{Dhar2015}. These methods alter specific frequency coefficients to encode watermark signals while ensuring minimal perceptual distortion. The work of \cite{hu2017incorporating} demonstrated a spectral-domain watermarking technique that modifies high-frequency bands to introduce secure identifiers without affecting speech intelligibility. Such methods exhibit strong robustness against common signal processing transformations, including MP3 compression and low-pass filtering, making them suitable for forensic analysis of AI-generated speech.

\textbf{Temporal-domain watermarking} embeds information within the time-domain features of an audio waveform, adjusting signal amplitude, phase, or timing characteristics. These techniques modify phoneme durations or introduce subtle phase shifts that remain undetectable to human listeners while ensuring verifiability through algorithmic analysis. One notable approach involves altering the pitch contour of AI-generated voices to encode a binary watermark signal while preserving natural speech prosody \citep{celik2005pitch}. Temporal watermarking methods tend to be more vulnerable to time-scale modifications such as speed adjustments and dynamic range compression, necessitating the use of hybrid techniques that integrate spectral and temporal features for enhanced robustness.

\textbf{Deep learning-based watermarking} represents an emerging paradigm that utilizes neural networks to learn optimal embedding strategies for speech watermarking. These methods train GANs or VAEs to encode watermark signals into the latent space of AI-generated speech, ensuring minimal perceptual distortion while maximizing robustness. AudioSeal \citep{roman:hal-04610152} exemplifies this approach, leveraging a convolutional neural network (CNN) to generate imperceptible watermark signals embedded within the spectrogram representation of speech. Deep learning-based speech watermarking offer superior adaptability to adversarial modifications but require substantial computational resources for both embedding and detection, limiting their real-time scalability.

\subsection{Watermarking for AI-Generated Music}

AI-generated music presents unique challenges for watermarking due to the complex interplay of melody, harmony, rhythm, and instrumentation. Watermarking techniques for AI-composed music must preserve musical integrity while ensuring resilience against transformations such as tempo changes, remixes, and dynamic equalization \citep{epple2024watermarking}. Music watermarking approaches are typically classified into symbolic-level and audio-level methods.

\textbf{Symbolic-level watermarking} operates on structured representations of music, such as MIDI files, embedding information within note sequences, rhythmic patterns, or harmonic progressions. These techniques introduce subtle modifications to chord voicings, note velocities, or timing variations to encode watermark signals while maintaining musical coherence. One prominent approach involves embedding unique harmonic transitions that remain detectable across different instrumentations and tempo adjustments \citep{bhandari2024motifs}. While highly robust within digital compositions, symbolic watermarking is ineffective for AI-generated music distributed as audio recordings.

\textbf{Audio-level watermarking} embeds watermarks directly into the audio waveform, leveraging spectral, phase, or amplitude modulation techniques. Frequency-domain watermarking is particularly effective, embedding signals in low-energy spectral bands that remain resilient to common post-processing transformations. Recent work by \cite{liu2023dear} explored phase-based watermarking methods that introduce imperceptible phase shifts across harmonic frequencies, ensuring robust detection while maintaining high audio fidelity. Audio-level watermarking techniques are widely applicable to commercial AI-generated music platforms but require careful parameter tuning to balance imperceptibility and detectability.

\textbf{Deep learning-driven watermarking} employs neural networks to optimize watermark embedding and extraction within AI-generated music. SynthID, developed by DeepMind, exemplifies this approach, integrating watermarking into the latent space of music generation models to ensure persistent traceability \citep{dathathri2024scalable}. These methods provide strong resilience against adversarial modifications but demand high computational overhead, making real-time deployment challenging. The integration of self-supervised learning techniques holds promise for improving the efficiency and adaptability of deep learning-based music watermarking \citep{singh2024silentcipher}.

\section{Challenges, Ethical Considerations, and Future Directions}
\label{sec_challenge_ethical_future}

Watermarking techniques for AI-generated content, while promising, encounter several persistent challenges that span across different modalities such as text, visual, and audio. Additionally, the implementation of watermarking raises significant ethical considerations that must be carefully navigated. This section delves into the primary challenges, ethical issues, and outlines potential future research directions essential for advancing watermarking methodologies.

\subsection{Robustness and Adversarial Attacks}

A paramount challenge in watermarking AI-generated content is ensuring robustness against adversarial attacks and transformation techniques. Watermarks embedded in text, visual, and audio content are susceptible to various forms of manipulation aimed at removing or obfuscating the embedded signatures. For instance, text watermarking techniques that rely on probabilistic token manipulation can be disrupted by paraphrasing or synonym substitution, as highlighted by \citep{sadasivan2023can}. Similarly, spatial-domain watermarking in visual content is vulnerable to common image processing operations such as compression, filtering, and geometric transformations \citep{sharma2024review}. In the audio domain, temporal-domain watermarking can be compromised by pitch shifting and dynamic range compression \citep{roman:hal-04610152}.

Deep learning-based watermarking approaches, while offering enhanced robustness, are not impervious to sophisticated adversarial strategies. Techniques such as GAN-based watermark removal and fine-tuning of diffusion models can effectively erase embedded watermarks without significantly degrading content quality \citep{sun2021detect}. Addressing these vulnerabilities requires the development of more resilient embedding strategies that can withstand a wide array of adversarial modifications. Hybrid watermarking methods that integrate spatial, frequency, and deep learning techniques may offer improved robustness by leveraging the strengths of multiple approaches \citep{singh2023comprehensive}.

\subsection{Standardization and Interoperability}

The lack of standardized watermarking protocols across different content modalities presents a significant barrier to the widespread adoption and interoperability of watermarking technologies. Interoperability, defined as the ability of different systems, technologies, or organizations to seamlessly exchange and effectively use information or functionalities without restrictions, is crucial for ensuring that watermarking solutions can function uniformly across diverse platforms and industries. Currently, watermarking schemes vary considerably between text, visual, and audio domains, each employing distinct embedding and detection methodologies \citep{zhao2024sok}. This heterogeneity complicates the establishment of universal benchmarks and hinders the ability to verify watermarks across diverse platforms and industries.

Establishing industry-wide standards is crucial to ensure consistency in watermarking practices and facilitate cross-platform verification. Standardization efforts should aim to define common evaluation metrics, embedding protocols, and verification procedures that can be uniformly applied across different types of content. Furthermore, the development of interoperable watermarking frameworks that can seamlessly integrate with various generative models is essential to improve the scalability and applicability of watermarking solutions \citep{simmons2024interoperable}.

\subsection{Ethical and Privacy Considerations}

The deployment of watermarking technologies raises several ethical and privacy concerns that must be addressed with care. One primary concern is the potential for unauthorized watermarking of human-generated content, which could infringe on individual privacy and content ownership rights. In the context of AI-generated speech, embedding metadata without user consent can lead to surveillance and tracking issues, posing significant ethical dilemmas \citep{findlay2025ai}.

Moreover, the voluntary adoption of watermarking by AI developers introduces challenges related to enforcement and compliance. Without regulatory mandates, the effectiveness of watermarking as a tool to verify the authenticity of content is undermined, as developers may opt to bypass watermarking mechanisms, especially in open-source models \citep{madiega2023generative}. Balancing the enforcement of watermarking practices with respect to digital rights and fair use is critical to ensuring that watermarking does not stifle innovation or infringe on user autonomy.

Ethical watermarking implementations should incorporate privacy-preserving techniques that allow for watermark verification without compromising the privacy of content creators and consumers. Techniques such as encrypted watermarking and key-based authentication can enhance security while safeguarding against unauthorized access and misuse \citep{cheng2022privacy}.

\subsection{Future Research Directions}

Future research in watermarking for AI-generated content should focus on developing more adaptive, resilient, and scalable methods that address evolving generative models and adversarial tactics. One promising direction is the advancement of adaptive watermarking techniques that dynamically adjust to changes in generative model architectures, ensuring robustness against evolving attacks \citep{liu2024adaptive}. Context-aware watermarking, which tailors embedding strategies based on domain-specific characteristics, can further enhance resilience \citep{guo2024context}.

Hybrid detection strategies that combine watermarking with retrieval-augmented verification and multimodal analysis hold significant potential for improving detection reliability. Retrieval-based detection leverages stored databases of AI-generated content to compare suspicious content against known watermarked outputs. By employing high-dimensional similarity searches and contrastive learning techniques, retrieval-based methods can effectively detect paraphrased or lightly edited AI-generated text, mitigating weaknesses in standalone watermarking schemes \citep{kang2024retrieval}. Similarly, retrieval-augmented approaches in visual and audio domains can enhance detection precision by cross-referencing content with pre-indexed AI-generated samples.

Ensuring watermarking robustness against adversarial attacks remains a pressing challenge. Generative models can be fine-tuned to erase embedded watermarks while maintaining content quality, necessitating watermarking schemes that resist transformations such as synonym substitution, recompression, and paraphrasing \citep{sadasivan2023can}. Techniques such as adversarially robust watermarking, which employs perturbation-resistant encoding strategies, offer a promising solution. Additionally, integrating self-supervised learning techniques can optimize the efficiency and adaptability of deep learning-based watermarking methods, facilitating real-time deployment and scalability \citep{singh2024digital}.

Standardization remains an essential research direction, as the lack of universal watermarking protocols hinders widespread adoption. Establishing industry-wide benchmarks and common evaluation metrics will improve interoperability between different watermarking techniques and content verification systems \citep{dathathri2024scalable}. Moreover, privacy-preserving watermarking techniques, such as encrypted and zero-watermarking schemes, are necessary to prevent unauthorized tracking while ensuring watermark detectability \citep{zhao2024sok}.

Finally, interdisciplinary research is crucial to bridging technical advancements with ethical, regulatory, and legal considerations. Regulatory frameworks must balance content authenticity verification with user rights, ensuring that watermarking does not infringe on privacy or creative freedoms. The integration of watermarking with broader content provenance initiatives, such as digital content authentication systems and blockchain-based verification, represents an important direction for future research.

\section{Conclusion}
\label{sec_conclusion}

Watermarking stands out as a crucial proactive mechanism for the detection and authentication of AI-generated content across text, visual, and audio modalities, addressing the escalating concerns of misinformation, identity fraud, and content manipulation inherent in the proliferation of GenAI technologies. This survey categorizes and evaluates the diverse watermarking methodologies, highlighting the delicate balance between imperceptibility, robustness, security, and embedding capacity that each technique must achieve to be effective. Despite notable progress in developing sophisticated watermarking strategies, significant challenges remain, including enhancing resilience against increasingly sophisticated adversarial attacks, achieving standardization across various content types to ensure interoperability, and addressing ethical and privacy implications associated with watermark deployment. Furthermore, the dynamic nature of generative models call for the continuous evolution of watermarking techniques to maintain their efficacy. Future research should prioritize the development of adaptive and hybrid watermarking approaches that leverage advancements in deep learning and cryptographic methods to improve robustness and security. Additionally, establishing industry-wide standards and best practices is essential for the widespread adoption of watermarking solutions. Addressing ethical concerns such as user privacy and content ownership is crucial for fostering trust and compliance. By overcoming these challenges, watermarking can maintain digital content integrity and ensure a trustworthy ecosystem in the era of GenAI.

\bibliography{iclr2025_conference}
\bibliographystyle{iclr2025_conference}

\end{document}